\documentclass[aps,prd]{revtex4}
\usepackage{graphics}
\begin{document} 

\title{
Constraining Warm Inflation with the Cosmic Microwave Background} 
\author{Lisa M. H. Hall}
\email{lisa.hall@ncl.ac.uk}
\author{Ian G. Moss}
\email{ian.moss@ncl.ac.uk}
\affiliation{School of Mathematics and Statistics, University of  
Newcastle Upon Tyne, NE1 7RU, UK}
\author{Arjun Berera} \email{ab@ph.ed.ac.uk}
\affiliation{School of Physics, University of Edinburgh,
Edinburgh, EH9 3JZ, U.K.}

\date{\today}

%%%%%%%%%%%%%%%%%%%%%%%%%%%%%%%%%%%%%%%%%%%%%%

\begin{abstract}

We discuss the spectrum of scalar density perturbations from warm inflation
when the friction coefficient $\Gamma$ in the inflaton equation is dependent
on the inflaton field. The spectral index of scalar fluctuations depends on a
new slow-roll parameter constructed from $\Gamma$. A numerical integration of
the perturbation equations is performed for a model of warm inflation and gives
a good fit to the WMAP data for reasonable values of the model's parameters.
\end{abstract}
\pacs{PACS number(s): 98.80.Cq, 98.80.-k, 98.80.Es}

\maketitle
%%%%%%%%%%%%%%%%%%%%%%%%%%%%%%%%%%%%%%%%%%%
%\section{introduction}

The inflationary paradigm has proved to be the most successful idea for
providing the seeds of large scale structure in the universe. A plethora of
variants of the inflationary model exist today and in most of these models,
radiation is red-shifted during the expansion, resulting in a vacuum-dominated
Universe, giving the name supercooled inflation. A subsequent reheating
period is invoked to end inflation and fill the universe with radiation
\cite{guth81,linde82,albrecht82,hawking82}.  In an alternative picture, termed
warm inflation, dissipative effects, arising from inflaton interactions, are
important during the inflation period, so that radiation production occurs
concurrently with inflationary expansion
\cite{moss85,yokoyama88,bererafang95,berera95,berera96,berera97}.

In this letter we shall demonstrate how it is possible to constrain a
particular type of warm inflationary model using Cosmic Microwave Background
observations. The model which we consider is based on broken supersymmetry
and has the property that the inflaton has a two stage decay chain
$\phi\to\chi\to\psi$ \cite{berera02,berera03}. This two stage decay occurs for
a wide range of inflationary models and provides a robust realisation of warm
inflation. In these models the friction coefficient $\Gamma$ in the inflaton
equation is a function of the inflaton field and this feature leads to some
new effects on the perturbation spectrum, which we describe here.

The development of cosmological perturbations in warm inflation has been
investigated previously by several authors. Taylor and Berera \cite{taylor00}
have analysed the perturbation spectra by identifying the small scale thermal
fluctuations with cosmological perturbations when the fluctuations cross the
horizon. This technique is good for order of magnitude estimates. 
For a more accurate treatment of the density perturbations it is necessary to
use the cosmological perturbation equations, which can be found in
\cite{lee00,oliveira01,hwang02}. A numerical code has recently
been developed which can solve the perturbation equations for any form of
friction coefficient \cite{hall03}.

We shall consider the case where the radiation produced by the inflaton field
consists of low mass particles $\psi$ which are approximately thermalised. 
In this case the density fluctuations
arise from thermal, rather than vacuum, fluctuations
\cite{moss85,bererafang95,berera95,berera96,berera00,hall03}.  The intermediate
particle $\chi$ is not in a thermal state and thermal corrections to the
inflaton potential are suppressed. This allows us to use the zero-temperature
potential during inflation. The effects of thermal corrections will be
discussed in future work.

Assuming the early universe to be in homogeneous expansion with expansion rate
$H(t)$, the evolution of the inflaton field $\phi$ is
described by the equation
\begin{equation}
\ddot\phi+(3H+\Gamma)\dot\phi+V_{,\phi}=0\label{wip}
\end{equation}
where $\Gamma(\phi)$ is the damping term and $V(\phi)$ is the
inflaton potential\cite{hosoya84,moss85,berera95}. The relative strength of the
thermal damping compared to the expansion damping can be expressed by a
parameter $r$,
\begin{equation}
r={\Gamma\over 3H}
\end{equation}
The strong dissipation form of warm inflation occurs for large values of $r$,
as opposed to supercooled inflation in which $r$ can be neglected.    

The dissipation of the inflaton's motion is associated with the production of
entropy. Energy-momentum conservation implies that
\begin{equation}
T(\dot s + 3 H s)= \Gamma\dot\phi^2.\label{wis}
\end{equation}
where $s$ is the entropy density. For the situation under discussion, the
entropy density is related to the energy density in the radiation $\rho_r$ by
$\rho_r=\frac34Ts$, where $T$ is the temperature, $s=\frac{\pi^2}{30}g_*T^3$
and $g_*$ is the effective number of degrees of freedom in the radiation. In
more general situations, using the entropy density is preferable to using the
radiation energy density \cite{hall03}.

The zero curvature Friedman equation completes the set of
differential
equations for $\phi$, $T$ and the scale factor $a$,
\begin{equation}
3H^2=8\pi G({\textstyle\frac12}\dot\phi^2+V+{\textstyle\frac34}Ts).\label{wia}
\end{equation}

The slow-roll approximation consists of neglecting terms
in the preceding equations with the highest order in time derivatives, 
\begin{eqnarray}
\dot\phi&=&{-V_{,\phi}\over 3H(1+r )}\label{slowp}\\
Ts&=& r \dot\phi^2\label{slows}\\
3H^2&=&8\pi GV.\label{slowh}
\end{eqnarray}
Slow-roll automatically implies inflation, $\ddot a>0$.

The consistency of the slow-roll approximation is governed
by a set of slow-roll parameters. Warm inflation has extra slow-roll
parameters in addition to the usual set for supercooled inflation due to the
presence
of the damping term $\Gamma$. The `leading order' slow-roll
parameters are,
\begin{equation}
\epsilon={1\over 16\pi G}\left({V_{,\phi}\over V}\right)^2,\quad
\eta={1\over 8\pi G}\left({V_{,\phi\phi}\over V}\right),\quad
\beta={1\over 8\pi G}\left({\Gamma_{,\phi}V_{,\phi}\over \Gamma V}\right)
\label{slowrp}
\end{equation}
The first two parameters are the standard ones introduced for supercooled
inflation \cite{liddle94,stewart02,habib02,martin03}. A new parameter is
required when there is $\phi$ dependence in the damping term. A further
slow-roll parameter is required when the potential has thermal corrections
\cite{hall03}. The slow-roll approximation is valid when all of these
parameters are smaller than $1+r$. In supercooled inflation, the
condition is tighter and the slow roll parameters have to be smaller than 1.

In the warm inflationary regime where $r\gg 1$, the
rate of change of various slowly varying parameters are given by
differentiating equations (\ref{slowp}-\ref{slowh}),
\begin{eqnarray}
{1\over H}{d\ln H\over dt}&=&-{1\over r}\epsilon\label{hdot}\\
{1\over H}{d\ln \dot\phi\over dt}&=&-{1\over r}(\eta-\beta)\\
{1\over H}{d\ln Ts \over dt}&=&-{1\over r}(2\eta-\beta-\epsilon)\label{tdot}
\end{eqnarray}
Inflation ends at the time when $\epsilon=r$, since then $\ddot a=\dot
H+H^2=0$. 

Cosmological perturbations are best described in terms of gauge invariant
quantities, the curvature perturbation ${\cal R}$ \cite{lukash80}, and the
entropy perturbation $e$ \cite{kodama84}. 
An analytic approximation to the density perturbation amplitude for wave number
$k$ can be obtained by matching the classical perturbations to the thermal
fluctuation amplitude at the
crossing time $k=aH$,
\begin{equation}
{\cal P}_{\cal R}\sim\left({\pi\over 4}\right)^{1/2}
{H^{5/2}\Gamma^{1/2}T\over\dot\phi^2}.\label{anr}
\end{equation}
This result is analogous to the result ${\cal P}_{\cal R}=H^4/\dot\phi^2$ for
supercooled inflation.

The spectral index $n_s$ is defined by
\begin{equation}
n_s-1={\partial \ln{\cal P}_{\cal R}\over\partial \ln k}.
\end{equation}
The slow-roll equations (\ref{hdot}-\ref{tdot}) enable us to express the
spectral index for the amplitude (\ref{anr}) in terms of slow-roll
parameters,
\begin{equation}
n_s-1={1\over r }\left(-{9\over 4}\epsilon+{3\over 2}\eta-
{9\over 4}\beta\right).\label{speci}
\end{equation}
The first two terms agree with reference \cite{berera00}. The $\beta$ term
shows the dependence of the spectrum on the gradient of the damping term.
For comparison, the spectral index for standard, or supercooled inflation, is
$n_s=1-6\epsilon+2\eta$ \cite{liddle92}.

There are two second-order slow-roll parameters
\begin{equation}
\zeta^2={1\over (8\pi G)^2}\left({V_{,\phi}V_{,\phi\phi\phi}\over
V^2}\right),\quad
\gamma={1\over 8\pi G}\left({\Gamma_{,\phi\phi}\over \Gamma}\right).
\label{soslowrp}
\end{equation}
With these we can obtain an expression for the slope of the spectral index,
\begin{equation}
{d n_s\over d\ln k}={1\over r^2
}\left(-\frac92\beta^2-\frac{27}4\epsilon^2-\frac92\epsilon\beta
+\frac{15}4\eta\beta+6\epsilon\eta-\frac32\zeta^2+\frac92\gamma\epsilon
\right).\label{runspeci}
\end{equation}
The first three terms form a negative definite combination, but other terms,
including $-\zeta^2$, can be positive.

The effect of $\beta$ in combination with `new inflation' potentials is
particularly interesting. Such potentials have a maximum at $\phi=0$
and a minimum at $\phi=\phi_0$ \cite{linde82,albrecht82,hawking82}. If we take
$\Gamma\propto\phi^b$, then the behaviour of
$n_s$ in the vicinity of the two extrema can be determined
\begin{equation}
n_s-1\to\cases{-|\eta|(2-3b)&$\phi\to 0$\cr
-\infty&$\phi\to \phi_0$\cr}.
\end{equation}
Therefore, if $b>2/3$, there will be a value $\phi_m$ at which $n_s=1$ and
$n_s$ is decreasing. The density perturbation amplitude has a maximum at the
wave number which crossed the horizon when $\phi=\phi_m$, and the spectrum
runs from blue to red as the wave number increases. The possibility of a
spectrum which runs from blue to red is particularly interesting because it is
not commonly seen in inflationary models, which typically predict red spectra.
(For examples of inflation with blue spectra, see \cite{copeland94,linde97}).

%\section{numerical results}

Now consider the spectrum of density perturbations for a model with
interaction Lagrangians $\frac12g^2\phi^2\chi^2$ and $h\chi\bar\psi\psi$. The
friction term can be approximated by \cite{berera02,berera03}
\begin{equation}
\Gamma\approx {g^3 h^2\phi\over 256\pi^2}.\label{brg}
\end{equation}
Warm inflation occurs when $g\approx h\approx 0.1$. For
coupling constants of this magnitude, quantum loop corrections to the inflaton
must be suppressed by an underlying supersymmetry to prevent them violating
the slow-roll conditions. The supersymmetry is broken during inflation,
either spontaneously, or by soft supersymmetry breaking masses. (These soft
terms are not necessarily the same soft terms which occur at the TeV scale). 

We take the soft supersymmetry breaking option, with inflaton potential,   
\begin{equation}
V(\phi)=\frac12\mu_0^2\left( \phi^2\log\left({\phi^2\over\phi_0^2}\right)
+\phi_0^2-\phi^2\right).
\end{equation}
The damping term $\Gamma$, will be parameterised by
\begin{equation}
\Gamma=\Gamma_0{\phi\over\phi_0}.\label{friction}
\end{equation}
With this combination, prolonged inflation (sixty e-folds or more)can occur for
values of $\phi_0$ smaller than the Planck scale $m_p$. The running of the
spectral index (\ref{runspeci}) is always negative. By contrast, in the
supercooled limit $\Gamma_0=0$, a prolonged period of inflation requires
$\phi_0$ to be at least $2m_p$ and it may not be consistent to ignore the
effects of quantum gravity. The density fluctuation amplitude for supercooled
inflation requires $\mu_0$ to be around $10^{13}$ GeV.

Our numerical code solves the homogeneous background and perturbation equations
simultaneously, with no assumption of slow-roll. The reader is referred to
\cite{hall03} for further details. The code solves for the primordial power
spectrum with the input parameters $\mu_0$, $\phi_0$ and $\Gamma_0$ defined
above. In order to compare to WMAP data, power law behaviour of the spectrum
is assumed about a scale $k_0=0.002 Mpc^{-1}$.  The power spectrum is then
characterised by the index, $n_s$ and spectral running $dn/d \ln k$.

We compute the microwave anisotropies using the CAMB code \cite{lewis99} and
run the WMAP likelihood code \cite{verde03} in order to compute the likelihood
fit of the model to WMAP data. Since at present we are interested in isolating
the behaviour of the above warm inflationary parameters, $\mu_0$, $\phi_0$ and
$\Gamma_0$, we assume priors for the cosmological parameters
($\Omega_\Lambda$, $\Omega_b$, $\Omega_m$ etc.).  The assumed parameter values
are given in table \ref{table1} and are taken from Bennett et al.\cite{map03}. 
The tensor components are assumed to be negligible.

\begin{table}[t]
\caption{\label{table1}Cosmological parameters taken from Bennett\cite{map03}. 
These priors are used as inputs into the CAMB code.}
\begin{tabular}{lcc}
Parameter & Symbol & Value \\ \hline
Dark energy density & $\Omega_\Lambda$ & 0.73 \\
Baryon density & $\Omega_b $ & 0.0044 \\
Matter density & $\Omega_m $ & 0.27 \\
Light neutrino density & $\Omega_\nu $ & 0 \\ 
Spatial curvature density & $\Omega_k $ & 0 \\
Hubble constant &h & 0.71 \\ \hline
\end{tabular}
\end{table}

For given values of $\mu_0$ and $\Gamma_0$, the value of $\phi_0$ is
chosen to normalise the amplitude of the power spectrum to the WMAP value
$A_{WMAP}$ at $k=k_0$. (In practice, we save some computation time by fixing
the value of $\phi_0$ and using renormalisation to rescale the model to the
correct amplitude.) We have restricted the range of $\phi_0$ to be less than
the Planck scale.

\begin{figure}[t]
\scalebox{0.5}{\includegraphics{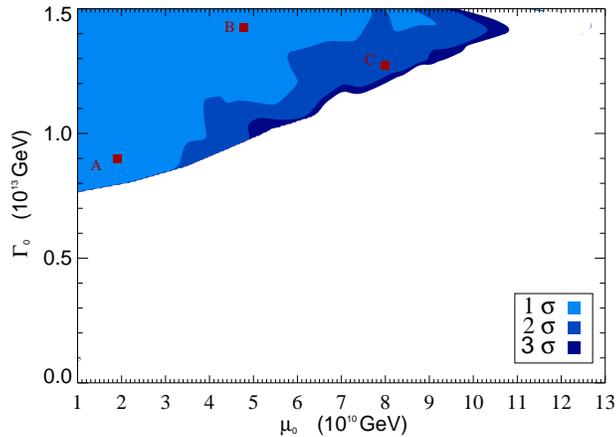}}
\caption{\label{fig1}The $1\sigma$, $2\sigma$ and $3\sigma$ bounds on the
parameters for the potential and friction term, $\mu_0$ and $\Gamma_0$ as
calculated with CAMB in conjunction with the WMAP likelihood code. Values for
the spectral index and spectral running at the points marked A, B and C are
given in table \ref{table2}.}
\end{figure}

Two inflationary parameters remain
to be investigated, namely $\mu_0$ and $\Gamma_0$. The $1\sigma$, $2\sigma$
and $3\sigma$ bounds for this parameter space are displayed in figure
\ref{fig1}. Note that there is no supercooled inflationary limit ($\Gamma_0
\rightarrow 0$) in figure \ref{fig1} because the density fluctuation amplitude
for supercooled inflation is too small in the parameter range shown.

The $3\sigma$ limit extends well beyond the upper edge of the figure $\Gamma_0
> 1.5 \times 10^{13} GeV$.  Consideration of the friction term (\ref{brg})
shows that a value of $\Gamma_0 = 1.5 \times 10^{13} GeV$ corresponds to
couplings, $g \approx h \approx 0.2$.  As $g$ and $h$ increase further still
($\Gamma_0$ increasing), we go beyond the perturbative regime in which the
potential which we are using is valid. We also expect thermal corrections to
the potential to be important when the coupling constants reach this
magnitude, and we shall return to this issue in future work.

\begin{table}[t]
\caption{\label{table2}Spectral index and running for points A, B and C in
figure \ref{fig1}}
\begin{tabular}{|l|c|c|} \hline 
           & {\bf $n_s$} & {\bf$d n_s~/~d \ln k$} \\ \hline 
{\bf A~~}  & 0.995 & $-2.1\times 10^{-4}$  \\ \hline
{\bf B~~}  &  0.994&  $-3.7\times 10^{-5}$ \\ \hline
{\bf C~~}  & 1.004 &  $-3.2\times 10^{-4}$\\ \hline
\end{tabular}
\end{table}

At points A, B and C in figure \ref{fig1}, the spectral index comes out
very close to $n_s=1$ with a very small amount of spectral running (see table
\ref{table2}). The running
of the spectrum is negative, as we predicted above, but rather small compared
to some values which have been discussed in connection with the WMAP data.

\begin{figure}[t]
\scalebox{0.7}{\includegraphics{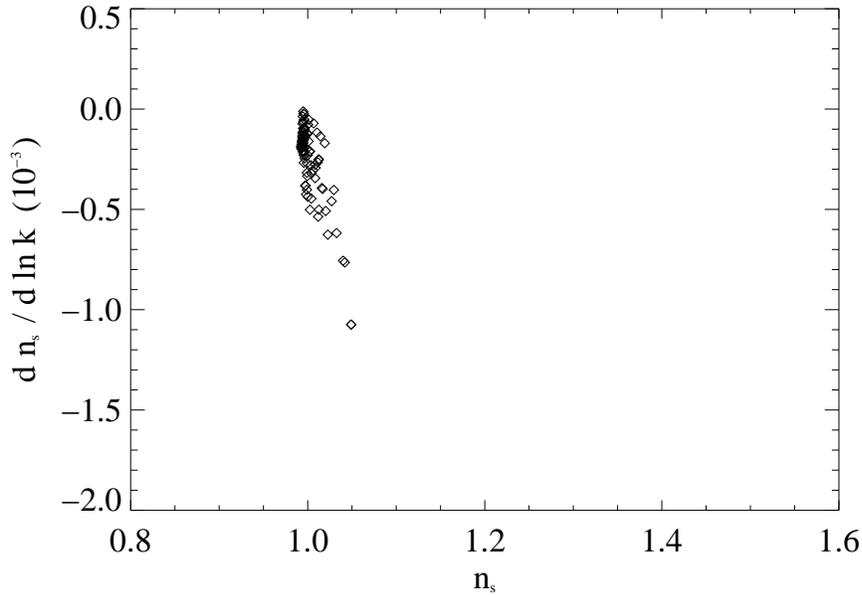}}
\caption{\label{fig2}These points represent the spectra produced for a
uniformly distributed selection of parameter values $\mu_0$ and $\Gamma_0$.
Note that the vertical axis is in units of $10^{-3}$.}
\end{figure}

Typical properties of the spectrum can be understood by analysing the
spectra produced by a uniformly distributed set of parameter values as in
figure \ref{fig2}. The parameter values are only included if inflation lasts
long enough to affect the large scale structure. The spectra produced by these
uniform parameter points are all within a small range around $n_s =1$ at a
scale $k_0=0.002 Mpc^{-1}$, with small spectral
running.  In contrast, the WMAP data can accommodate large spectral running if
there are relatively large departures from $n_s =1$ \cite{map03-2}, and the
observations are not yet accurate enough to provide very strong restrictions on
the parameter values.

%\section{conclusion}

We conclude by summing up the main results.  For warm inflation, a third
slow-roll parameter is introduced due to the dependence of the spectrum on the
dissipation term, $\Gamma$.  For inflation with new inflation potentials, it
has been shown that, in warm inflation, one expects the spectral index to run
from blue to red as the wavenumber increases.  This behaviour is particularly
interesting in the light of the first year WMAP data, in which such a running
is hinted at.  For a specific potential, representative of supersymmetry
breaking, warm inflation has been compared to the WMAP data and a good fit was
found for a large range of parameter values. The model produces spectra with
$n_s$ close to one and a very small negative running of the index. Future
observations of the Cosmic Microwave Background are needed to detect such
small slopes in the spectrum and produce stronger constraints on the
parameters.

\acknowledgements{LH thanks Sujata Gupta, David Parkinson and Sam Leach for
many useful discussions.}

%%%%%%%%%%%%%%%%%%%%%%%%%%%%%%%%%%%%%%%%%%%%%
\bibliography{lisa.bib}

%%%%%%%%%%%%%%%%%%%%%%%%%%%%%%%%%%%%%%%%%%%%%%
\end{document}